\documentclass[fleqn,12pt,twoside]{article}
\usepackage{espcrc1}
\usepackage{epsfig}
\begin{document}
\title{Symmetries and Symmetry Breaking
\footnote{This work was supported in part by the Natural Sciences and Engineering
Research Council of Canada}}

\author{Willem T.H. van Oers\\
TRIUMF, 4004 Wesbrook Mall, Vancouver, BC Canada V7H 2Y6\\
and\\
Department of Physics and Astronomy, University of Manitoba,
  Winnipeg, MB, R3T 2N2, Canada}

\maketitle


In understanding the world of matter the introduction of symmetry
principles following experimentation or using the predictive
power of symmetry principles to guide experimentation is most
profound. The conservation of energy, linear momentum, angular
momentum, charge, and CPT involve fundamental symmetries. All other
conservation laws are valid within a restricted subspace of the four
interactions: the strong, the electromagnetic, the weak, and the
gravitational interaction.
In what follows comments will be made regarding parity violation in
hadronic systems, charge symmetry breaking in two nucleon and few
nucleon systems, and time-reversal-invariance in hadronic systems.

In hadronic parity violation, theory and experiment meet in the weak
meson-nucleon couplings as defined, for instance, in the seminal article
by Desplanques, Donoghue, and Holstein (DDH). [1] In analogy with the one
boson exchange model for the strong N-N interaction, one defines seven
weak meson-nucleon couplings for $\pi$, $\rho$, and $\omega$ exchanges and
according the change in isospin. These seven weak meson-nucleon couplings
or another set of seven parameters in another theoretical framework
characterize the hadronic weak interaction. For a more recent calculation
of the weak meson-nucleon couplings see Feldman, Crawford, Dubach, and
Holstein. [2] The latest experimental contribution to the determination
of the weak meson-nucleon couplings is the measurement at TRIUMF of the
longitudinal analyzing power, $A_z~=~(\sigma^{+}~-~\sigma^{-})/(\sigma^{+}~
+~\sigma^{-})$, in polarized proton-proton scattering at 221 MeV,
where $\sigma^{+}$ and $\sigma^{-}$ are the scattering cross sections
for positive and negative helicity, respectively. The measurements were
performed in transmission geometry with the beam energy and the detector
layout selected to ensure that the contribution of the lowest order parity
mixing amplitude $(^{1}S_{0}~-~^{3}P_{0})$ integrates to zero, hence
leaving essentially only a contribution of the $(^{3}P_{2}~-~^{1}D_{2})$
parity mixing amplitude. As a consequence the experiment determined the
weak meson-nucleon coupling $h^{pp}_\rho$. In the experiment the beam
current was measured before and after passing a 0.40 m thick liquid
hydrogen target by transverse electric field ionization chambers (TRICs),
and the difference determined as function of the helicity
state of the incident proton beam. Since many beam parameters may change
upon helicity flips and be the cause of false asymmetries, great care was
taken to have the sensitivity of the data taking apparatus to the unwanted
beam parameters minimized as much as possible, and further to minimize
the presence of the unwanted beam parameters, and finally to make
corrections to the data where necessary. Sets of diagnostic, monitoring,
and control instrumentation preceded the actual parity data taking
apparatus: polarization profile monitors (PPMs) and intensity profile
monitors (IPMs), ferrite cored steering magnets, as well as an energy
modulating RF cavity. The polarization
profile monitors measured on line the unwanted transverse polarization
components, $P_x$ and $P_y$, and their first moments, $yP_x$ and $xP_y$,
across the profile of the beam. The intensity profile monitors measured
similarly the intensity across the beam horizontally (x) and
vertically (y). The latter allowed to keep the beam position fixed in
two places upstream of the parity data taking apparatus. An integral part of
the experiment was the optically pumped polarized ion source (OPPIS).
Great efforts were made to suppress unwanted helicity correlated changes
in the polarized beam extracted from OPPIS. The final result of the 221 MeV
experiment is $A_z~=~(0.84~\pm~0.29(stat.)~\pm~0.17(syst.))~\times~10^{-7})$.
[3] Together with the low-energy proton-proton data at 13.6 MeV [4] and at
45 MeV [5], which determined a linear combination of $h^{pp}_\rho$ and
$h^{pp}_\omega$, for the first time strong constraints could be placed on
the acceptable values of both $h^{pp}_\rho$ and $h^{pp}_\omega$. Many
theoretical calculations of $A_z$ have been made. Representative
calculations are depicted in Fig.1 together with the TRIUMF result
(corrected to the $(^{1}S_{0}~-~^{3}P_{0})$ zero crossing energy) and the
most precise low energy results at 13.6 MeV and 45 MeV. The prediction
of Driscoll and Miller [6] is based on the Bonn potential to represent
the strong N-N interaction, together with the weak meson-nucleon couplings
from Desplanques, Donoghue, and Holstein. The prediction of Iqbal and
Niskanen [7] has a $\Delta$ isobar contribution added to the Driscoll and
Miller theoretical model on a semi ad-hoc basis. The prediction of Driscoll
and Meissner [8] is based on a self-consistent theoretical model, with
both weak and strong vertex functions obtained with a chiral soliton
approach. Finally, the quark model prediction of Grach and Shmatikov [9]
takes explicit account of quark degrees of freedom. None of these
predictions are in good agreement with the data, although they have similar
shapes due to the energy dependence of the strong interaction phase shifts.

\begin{figure}[t]
\begin{center}
\vspace*{-4mm}
\epsfig{figure=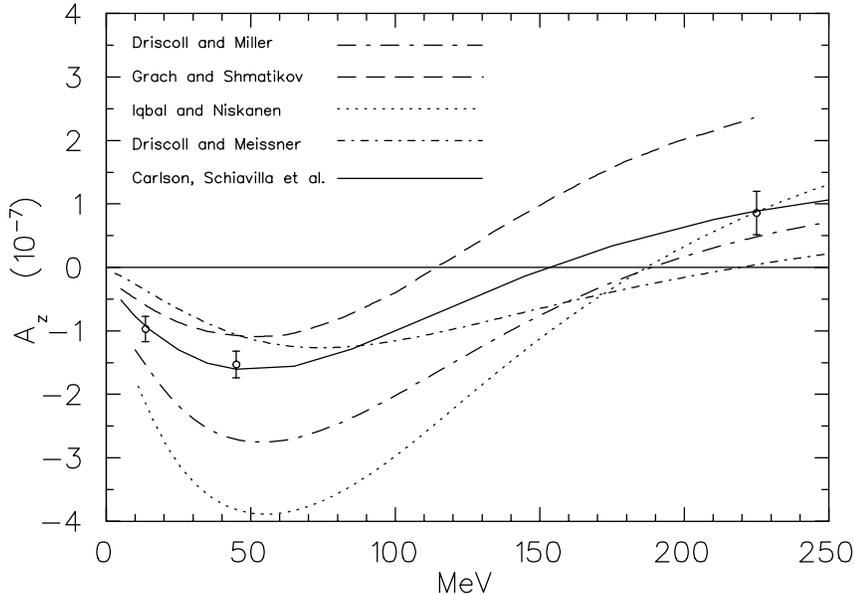,width=.70\linewidth}
\vspace*{-8mm}
\end{center}
\caption{Theoretical predictions for $A_z$ and the more precise data
at 13.6 MeV (Bonn), 45 MeV (PSI), and 221 MeV (TRIUMF). The solid curve
shows the result obtained by Carlson \emph{et al.} by adjusting the weak
meson-nucleon couplings for the best fit to the data.}
\label{data}
\vspace*{-4mm}
\end{figure}

The limits on the weak meson-nucleon couplings $h^{pp}_\rho$ and
$h^{pp}_\omega$ imposed by the low energy results at 13.6 and 45 MeV and
by the TRIUMF result at 221 MeV are shown in Fig.2. The error bands
$(\pm~1~\sigma)$ are based on a calculation by Carlson \emph{et al.} [10] assuming
the Argonne $v_{18}$ (AV-18) N-N potential, the Bonn 2000 (CD-Bonn) strong
interaction coupling constants, and including all partial waves up to
$J~=~8$. Although the TRIUMF measurement does not have a contribution from
the $(^{1}S_{0}~-~^{3}P_{0})$ parity violating mixing amplitude, and the
$(^{3}P_{2}~-~^{1}D_{2})$ parity violating mixing amplitude does not
contain $h^{pp}_\omega$, there is some $h^{pp}_\omega$ dependence arising
from the higher partial wave mixing amplitudes, i.e.,
$(^{1}D_{2}~-~^{3}F_{2})$. As a result the acceptable band as determined
by the TRIUMF experiment is almost perpendicular to those determined by
the low energy measurements and greatly reduces the acceptable ranges of
both $h^{pp}_\rho$ and $h^{pp}_\omega$. Adjusting these couplings for
the best fit to the three data points, Carlson \emph{et al.} estimate
$h^{pp}_\rho~=~-22.3~\times~10^{-7}$ and
  $h^{pp}_\omega~=~5.17~\times~10^{-7}$ compared to the DDH ``best guess''
values of $h^{pp}_\rho~=~-15.5~\times~10^{-7}$ and
$h{pp}_\omega~=~-3.0~\times~10^{-7}$. The solid curve in Fig.1 is calculated
using the adjusted couplings.

\begin{figure}[tbh]
\begin{center}
\vspace*{-4mm}
\epsfig{figure=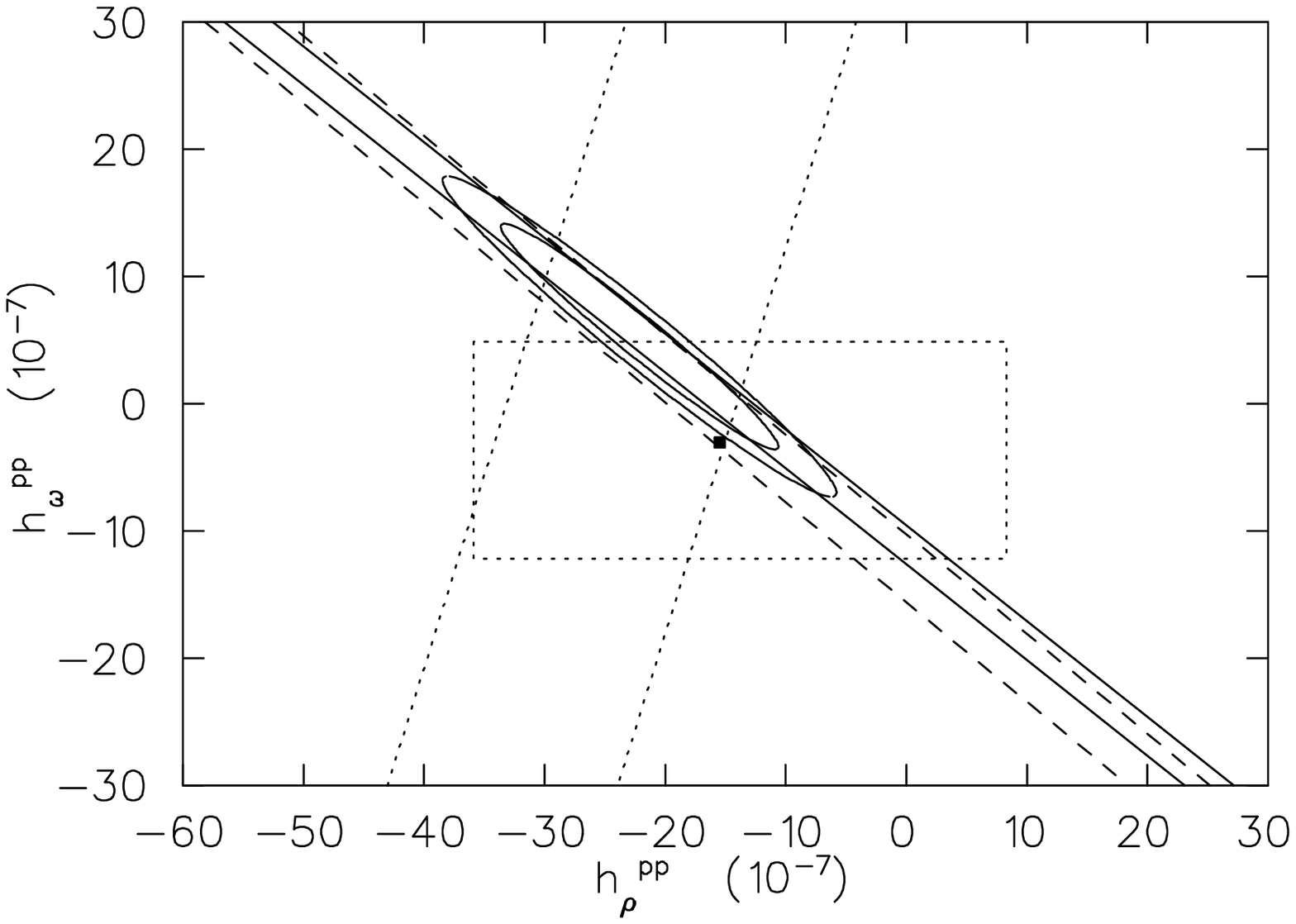,width=.70\linewidth}
\vspace*{-6mm}
\end{center}
\caption{{Present constraints on the weak meson-nucleon couplings based on
the experimental data and recent calculations by Carlson \emph{et al.}. The bands are
the constraints imposed by the different experiments (13.6 MeV Bonn, dashed;
45 MeV PSI, solid; 221 MeV TRIUMF, dotted. The filled square and dotted
rectangle are the DDH ``best guess'' and ``reasonable range'', respectively.
Also shown are the $68\%$ and $90\%$ C.L. contours.}}
\label{constraints}
\vspace*{-8mm}
\end{figure}

The current situation regarding the long range weak meson-nucleon coupling
$h^{1}_\pi$ is not as satisfactory, with large discrepancies between the
value extracted from the parity mixed doublet in $^{18}F$ in a series of
independent experiments and from atomic parity violation in Cs and Tl.
A new precision experiment has been mounted at LANL to measure
parity violation in polarized cold neutron capture on para-hydrogen. [11]
A precise determination of the seven weak meson-nucleon couplings is a
prerequisite to verify the many steps of imprinting the weak interaction
among mainly strong interacting quarks and gluons. Further
experimentation measuring spin rotation of cold neutrons through helium
and hydrogen, measuring the asymmetry in photodisintegration of the
deuteron with left-handed and right-handed circularly polarized photons,
and measuring the longitudinal analyzing power in polarized proton-proton
scattering at 221 MeV with greatly improved precision are essential to
precisely determine the seven weak meson-nucleon couplings and
to understand the intricacies of hadronic parity violation. A greatly
improved 221 MeV proton-proton parity violation experiment is possible
after adding current readout to the counting polarimeters (PPMs),
by replacing the transverse electric field ionization chambers (TRICs)
by current monitors, and by changing the analog electronics with
state-of-the-art digital electronics.

Isospin conservation/charge independence is broken by the mass difference
of the 'up' and 'down' quarks and by their electromagnetic interaction.
Charge symmetry is a lesser symmetry and corresponds to a rotation in
isospin space around the 2-axis. For the neutron-proton system one
classifies the interaction as an isospin breaking, charge-asymmetric,
and charge-dependent force. Charge symmetry breaking in neutron-proton
elastic scattering, manifested through a non-zero value of the difference
in the analyzing powers for polarized (unpolarized) neutrons scattered
from unpolarized (polarized) protons, is well established. The results of
two new charge symmetry breaking experiments will be reported at this
Conference. The first one (a TRIUMF experiment) has measured the
forward/backward asymmetry in the reaction $n-p~\rightarrow~d-\pi^{0}$
close to threshold (with a little less than a $2~\sigma$ effect)
($A_{fb}~=~(1.7~\pm~0.55(stat.)~\pm~0.8(syst.))~\times~10^{-3}$). [12] In
terms of meson-exchange nucleon-nucleon interaction models, the chief
contributions to charge symmetry breaking in this reaction are
$\pi~-~\eta$ mixing and the 'up' and 'down' quark mass difference affecting
pion-nucleon scattering. The second one (a IUCF experiment) has measured
the reaction $d-d~\rightarrow~\alpha-\pi^{0}$ just above threshold for
$\pi^{0}$ production. Evidence presented for $\pi^{0}$ production in this
reaction in an earlier experiment has been controversial. Employing a
three-fold coincidence between the $\alpha$-particle and the two
$\gamma$'s from the decay of the $\pi^{0}$, with careful reconstruction
of the mass of the latter, determined unambiguously a small cross section
for this charge symmetry breaking reaction, in accord with estimates based
predominantly on 'up' and 'down' quark mass difference effects
($\sigma_{tot}(228.5~MeV)~=~12.7~\pm~2.2~pb$ and $\sigma_{tot}(231.8~MeV)~=~
15.1~\pm~3.1~pb$). [13] These charge symmetry breaking experiments test the
details of meson-exchange nucleon-nucleon interaction models and delineate
the contributions from the mass difference of the up and down quarks and
their electromagnetic interaction.

Charge symmetry breaking in neutron-proton elastic scattering, mentioned
above, also permits an upper limit to be placed on a parity-conserving/
time-reversal-invariance-nonconserving nucleon-nucleon interaction
mediated by $\rho$ and $a_{1}-meson$ exchanges. [14] To greatly decrease
the current upper limit on this interaction requires a vastly improved
charge symmetry breaking neutron-proton elastic scattering experiment at
an energy of 320 MeV. At this energy and at the zero-crossing angle of
the neutron-proton analyzing power, where $A_z$ is measured, the
contribution from the uncertain $\rho~-~\omega$ mixing term is very small.
The electromagnetic contribution as well as the neutron proton mass
difference affecting $\pi$ and $\rho$ exchanges can be calculated with
confidence. Thus from a comparison between experiment and theory one can
deduce an improved upper limit on a parity-even/time-reversal-odd
interaction. From an experimental point of view, however, one prefers a
genuine null test of time-reversal-invariance, which is only possible in
total cross section measurements (attenuation experiments). Such an
experiment is being prepared for execution at COSY. [15] The experiment
will measure the parity-even/time-reversal-odd observable $A_{y,xz}$ in
transmission geometry. The experiment will use a stored, transversely
polarized proton beam (with polarization $P_{y}$) and an aligned polarized
deuterium target (with polarization $P_{xz}$). The tensor polarized deuteron
beam is produced in an atomic beam source based on Stern-Gerlach separation
in permanent sextupole magnets and adiabetic high frequency transitions.
Adequate luminosity is obtained using a windowless storage cell placed on
axis of the polarized proton beam from the atomic crossed-beam polarized
ion source. For this test of time-reversal-invariance the COSY-ring will
function as accelerator, forward angle spectrometer, and detector. Crucial
to the experiment is a current monitor, which can measure with the required
high precision the decrease in intensity with time as function of the
circulating proton beam spin state or the direction of the deuteron tensor
polarization. Also of great importance is the precise
alignment of the proton beam vector and deuteron beam tensor polarizations
in order to suppress unwanted spin correlation coefficients, which could
produce a false result. The stated precision of the experiment is one part
in $10^{6}$ for $A_{y,xz}$, which would give a sensitivity to
$\bar{g}_\rho$ of $10^{-3}$ by appropriate choice of the proton beam energy
($\bar{g}_\rho$ is the time-reversal-invariance nonconserving fraction of
the strong $\rho$-meson-nucleon coupling constant). This is a similar
upper limit as deduced from the neutron-proton elastic scattering charge
symmetry breaking experiments mentioned above. The experiment is currently
in an engineering phase. It should be noted that these are the better
upper limits on a parity-even/time-reversal-odd interaction in hadronic
systems that exist to date. But it should also be noted that there is no
room for such an interaction within the Standard Model; it would require
flavor non-conservation of the quarks.

\end{document}